\documentclass[12pt,preprint]{aastex}

\newcommand{\msun}{\mbox{ M$_\sun$}}
\newcommand{\bq}{\begin{equation}}
\newcommand{\eq}{\end{equation}}
\newcommand{\days}{\mbox{ days}}

\newcommand{\mjy}{\mbox{ mJy}}

\newcommand{\mhz}{\mbox{ MHz}}
\newcommand{\ghz}{\mbox{ GHz}}
\newcommand{\khz}{\mbox{ kHz}}

\newcommand{\kel}{\mbox{ K}}

\newcommand{\cm}{\mbox{ cm}}
\newcommand{\secinv}{\mbox{ s$^{-1}$}}
\newcommand{\aefftsys}{\mbox{ m$^2$  K$^{-1}$}}
\newcommand{\fluxden}{\mbox{ erg cm$^{-2}$  s$^{-1}$ Hz$^{-1}$ sr$^{-1}$}} 

\newcommand{\hunits}{\mbox{ km s$^{-1}$ Mpc$^{-1}$}}
\begin{document}
 
\title{The 21 cm Forest: Radio Absorption Spectra as Probes of
Minihalos Before Reionization} 

\author{Steven R. Furlanetto \& Abraham Loeb}
\affil{Harvard-Smithsonian Center for Astrophysics, 60 Garden St.,
Cambridge, MA 02138;\\
sfurlanetto@cfa.harvard.edu, aloeb@cfa.harvard.edu}

\begin{abstract}

We study the absorption along lines of sight toward high-$z$ radio
sources caused by the $21 \cm$ transition of neutral hydrogen in the
intergalactic medium (IGM) before reionization.  Using semi-analytic
methods, we compute the number density of observable features caused
by both ``minihalos'' (bound objects that are unable to cool
efficiently because of their small virial temperatures) and
protogalactic disks.  We show that both sets of features should be
observable by the next generation of low-frequency radio telescopes,
including the \emph{Low Frequency Array} and the \emph{Square
Kilometer Array}, provided that sufficiently bright background sources
exist.  The statistics of minihalo absorption features seen along
lines of sight to radio-loud quasars offer a way to measure the
evolution of the radiation background and the IGM temperature with
cosmic time.  Intersections with disks are much less common but cause
significantly deeper absorption features that would be visible in the
spectra of both radio-loud quasars and gamma-ray bursts (GRBs).  The
absorption feature caused by HI in the host galaxy of a GRB should be
observable, offering a route to determine spectroscopically the burst
redshift.

\end{abstract}


\section{ Introduction }
\label{intro}

Despite its apparent simplicity, effective observational probes of the
intergalactic medium (IGM) at high redshifts are difficult to find.
Between recombination at $z \sim 1000$ and reionization at $z \sim 6$,
the IGM was almost entirely neutral.  In such a medium, the optical
depth to Ly$\alpha$ absorption at the frequency corresponding to a
redshift $z$ is $\tau_\alpha \approx 6.45 \times 10^5 x_{\rm HI}
[(1+z)/10]^{3/2}$ \citep{gp}, where $x_{\rm HI}$ is the neutral
fraction and where we have assumed the currently favored cosmological
parameters (see below).  This enormous optical depth renders the
Ly$\alpha$ forest, which is the most important tool for studying the
(highly ionized) IGM at low and moderate redshifts, almost useless for
detailed studies of the pre-reionization IGM.  For example, the
detection of a complete absorption trough shortward of the Ly$\alpha$
resonance wavelength in the spectrum of the quasar SDSS 1030+0524 may
be the first evidence for the end of the reionization era
\citep{becker}.  However, studies of rest-frame UV and optical quasar
spectra are unable to place stronger constraints on the neutral
fraction than $x_{\rm HI} \ga 10^{-2}$ \citep{fan}.  Attempts to
constrain other parameters, such as the thermal state of the IGM, the
radiation background, or the density structure, are also compromised
by the extremely large optical depth at the Ly$\alpha$ resonance.

Clearly, we cannot obtain an in-depth understanding of the IGM before
and during reionization through studies of the Ly$\alpha$ forest.  An
alternative is to observe a weaker transition at the other extreme of
optical depth: the $21 \cm$ hyperfine transition of neutral hydrogen.
To date, there have been several theoretical studies of emission or
absorption of the neutral IGM against the cosmic microwave background
(CMB) in this transition \citep{scott,kumar,mmr,tozzi,shaver,iliev}.
However, the predicted signals are extremely weak.  Even the next
generation of radio telescopes, such as the \emph{Square Kilometer
Array}\footnote{See, e.g., http://www.usska.org/main.html.} (SKA),
will only be able to detect exceptionally large objects (e.g.,
\citealt{scott,kumar}).  Another strategy is to study the angular
structure of the emission in order to constrain statistically the
structure formation process at high redshift (e.g.,
\citealt{tozzi,iliev}).  However, source confusion with low-$z$ faint
radio sources \citep{dimatteo} and free-free emission by the
intergalactic medium (Loeb 1996) are likely to compromise such
attempts.

Alternatively, \citet{carilli} have pointed out that luminous
high-redshift radio-loud quasars serve as ideal background sources
against which absorption by intervening gas can be seen: with a bright
background source it becomes much easier to identify the weak
absorption features expected to be produced by the IGM.  Along such
lines of sight one can map the ``21 cm forest'' of redshifted
hyperfine absorption lines and hence study the neutral IGM over the
redshift range $6 \la z \la 10$.  Using a simulation, \citet{carilli}
found that the high-$z$ analog of the Ly$\alpha$ forest of filaments
and other overdense regions produces recognizable absorption features
in the spectrum of a radio-loud quasar.  They also demonstrated that
such observations are feasible with an instrument similar to current
designs for the SKA, provided that sufficiently bright radio-loud
quasars exist at these early epochs.

Large scale numerical simulations, such as those studied by
\citet{carilli}, are unable to resolve collapsed gas clouds on the
smallest mass scales, although these clouds represent the most
abundant clumps in the high-redshift IGM.  In this paper we complement
their study by exploring semi-analytically the statistics of these
compact absorbing systems. Our semi-analytic approach allows us to
examine the sensitivity of the results to changes in the input
parameters and to identify the physical quantities that are best
measured by radio absorption spectra.  As we demonstrate in the
following sections, 21 cm absorption spectra probe the radiation
background and thermal state of the IGM at high redshifts and
therefore could be instrumental in testing models of structure
formation and reionization.

The smallest bound objects in the IGM (commonly referred to as 
``minihalos'') have virial temperatures below the threshold for
atomic hydrogen line cooling, and so (in the absence of a substantial
abundance of molecular hydrogen, H$_2$) they cannot cool and collapse
to form protogalaxies.  Although minihalos cannot form stars, they are
important for determining the mean clumping factor of the IGM and for
screening ionizing radiation from other objects \citep{barkana02}.
Their properties are sensitive to the uncertain presence of molecular
hydrogen; if H$_2$ forms in sufficient quantities it can act as a
cooling channel for these minihalos (\citealt{barkana01}, and
references therein), strongly suppressing their number density (and
hence increasing the global star formation rate).  Detection of these
halos would provide important insights into the physics of
hierarchical structure formation prior to reionization.

Halos with virial temperatures above the hydrogen line cooling
threshold will collapse to form stars.  Thus we expect some fraction
of all lines of sight to penetrate protogalactic disks.  Such
absorption systems would be the pre-reionization analogs of damped
Ly$\alpha$ absorbers.  While disk intersections are rare, they can
provide important information about the state of the neutral hydrogen
in high-$z$ galaxies, including the distribution of disk masses, star
formation rates, and gas temperature.

In \S \ref{model} we derive the optical depth to redshifted $21 \cm$
radiation through both the diffuse IGM and collapsed objects.  We show
our numerical results in \S \ref{results} and conclude in \S
\ref{disc}.  Throughout the discussion we assume a flat,
$\Lambda$--dominated cosmology, with density parameters $\Omega_m =
0.3$, $\Omega_\Lambda = 0.7$, $\Omega_b = 0.05$, in matter,
cosmological constant, and baryons, respectively.  In the numerical
calculations, we assume $h=0.7$, where the Hubble constant is $H_0 =
100 h \hunits$.

\section{ Model }
\label{model}

\subsection{ The Diffuse IGM }

The optical depth of the neutral IGM to redshifted hyperfine
absorption is \citep{field59a,mmr} 
\begin{eqnarray}
\tau_\nu & = & \frac{ 3 c^3 h_P A_{10} \, n_{HI}(z)}{32 \pi
k \nu_0^2 \, T_S \, H(z) } \nonumber \\
\, & \approx & 10^{-2} \left[ \frac{T_{\rm CMB}(z)}{T_S} \right]
\left( \frac{\Omega_b h}{0.035} \right) \left[
\left(\frac{0.3}{\Omega_m} \right) \, \left(
\frac{1+z}{10} \right) \right]^{1/2} x_{\rm HI}.
\label{eq:tauigm}
\end{eqnarray}
Here $h_P$ is Planck's constant, $k$ is Boltzmann's constant,
$\nu_0=1420.4 \mhz=\nu/(1+z)$ is the rest-frame hyperfine transition
frequency, $A_{10} = 2.85 \times 10^{-15} \secinv$ is the spontaneous
emission coefficient for the transition, $T_S$ is the spin temperature
of the IGM, $T_{\rm CMB} = 2.73 (1+z) \kel$ is the CMB temperature at
redshift $z$, and $n_{HI}$ is the neutral hydrogen density at this
redshift.  In the second equality, we have assumed sufficiently high
redshifts such that $H(z) \approx H_0 \Omega_m^{1/2} (1+z)^{3/2}$
(which is well-satisfied in the era we study, $z_r \ga 6$).  Because
reionization is expected to be sudden (e.g., \citealt{gnedin00}) we take
a neutral fraction $x_{\rm HI} = 1$ throughout our discussion.

The HI spin temperature (i.e., the excitation temperature of the
hyperfine transition) is \citep{field58}
\bq
T_S = \frac{T_{\rm CMB} + y_\alpha T_\alpha + y_c T_K}{1 + y_\alpha +
y_c}. 
\label{eq:hItspin}
\eq 
Here $T_K$ is the HI kinetic temperature and $T_\alpha$ is the
Ly$\alpha$ color temperature (see \citealt{mmr} for a precise
definition).  In the conditions pertaining to both the IGM and
collapsed objects, $T_\alpha = T_K$ \citep{field59b}.  The third term
describes collisional excitation of $T_S$, while the second describes
the Wouthuysen-Field effect, in which Ly$\alpha$ pumping couples the
spin temperature to the temperature of the radiation field
\citep{wout,field58}.  The coupling from collisions is 
\bq 
y_c = \frac{C_{10}}{A_{10}} \,
\frac{T_\star}{T_K},
\label{eq:yc}
\eq
where $C_{10}$ is the collisional de-excitation rate of the
(higher-energy) triplet
hyperfine level and $T_\star = h_P \nu_0/k = 0.068 \kel$.  We use a fit
to the results of \citet{allison} for $C_{10}$ at $T_K < 1000 \kel$,
with an extrapolation $C_{10} \propto T_K^{-0.33}$ to higher
temperatures.  The Wouthuysen-Field coupling constant is
\bq
y_\alpha = \frac{P_{10}}{A_{10}} \, \frac{T_\star}{T_\alpha},
\label{eq:yalpha}
\eq 
where $P_{10}$ is the indirect de-excitation rate of the triplet
level due to absorption of a Ly$\alpha$ photon followed by decay to the
singlet level.  For a diffuse Ly$\alpha$ background, \citet{mmr}
showed that 
\bq 
P_{10} \approx 1.3 \times 10^{-12} J_{-21} \secinv,
\label{eq:j21}
\eq 
where $J_{-21}$ is the intensity of the background radiation field
at the Ly$\alpha$ frequency in units of $10^{-21} \fluxden$.  Within
the diffuse IGM, the gas density is small enough that collisions do not
strongly couple $T_S$ and $T_K$.  However, collisions can effectively
couple the two quantities within collapsed objects.  Ly$\alpha$
pumping couples the two temperatures in a density-independent manner,
becoming effective when $J_{-21} \ga 1$.  Below we show results for
several values of $J_{-21}$, because the UV radiation field before
reionization is an unknown quantity, likely varying strongly with both
position and time.  For reference, the simulation used by
\citet{carilli} yields $J_{-21}(z=10) \approx 0.8$ and $J_{-21}(z=8)
\approx 4$.

The spin temperature and optical depth therefore depend on the kinetic
temperature of the IGM.  Once Thomson scattering with the CMB becomes
inefficient at the thermal decoupling redshift $z_d \sim 140$, the IGM
cools adiabatically as the universe expands \citep{barkana01} until
the first objects collapse.  As soon as this occurs, radiative (and
mechanical) feedback from these objects quickly heats the IGM.
Because the radiation background prior to reionization is still
unconstrained by observations, estimates of the temperature rely on
simulations.  We choose a particular form for the temperature
evolution that roughly matches the mass-weighted temperature found in
the simulation of \citet{carilli}: 
\bq 
T_{\rm IGM} (z) = 10^{-0.25 z + 5.5} \kel.
\label{eq:Tigm}
\eq 
This simple form has two convenient properties.  First, heating
begins at $z \sim 18$, close to the time when the first stars are
expected to form.  Second, it yields $T_{\rm IGM}(z=6) = 10^4 \kel$,
close to the temperature of a photo-ionized medium at approximately
the time when reionization may have occurred.  We neglect any spatial
variation in the temperature field, as is likely to occur in filaments
and other overdense regions.  For contrast, we include below some
results for zero heating and for a temperature that is smaller than
that given in equation (\ref{eq:Tigm}) by a factor of $10$.

\subsection{ Minihalos }

In cold dark matter models, structure forms hierarchically (i.e.,
low-mass halos collapse first and the characteristic mass scale
increases with cosmic time).  The smallest halos cannot, however, form
stars because they cannot cool efficiently: without metals, atomic
hydrogen line cooling is only effective for halos with a virial
temperature $T_{\rm vir} \ga 10^4 \kel$.  Here, $T_{\rm vir}$ is
defined as in equation (26) of \citet{barkana01}.  We use the term
\emph{minihalo} to refer to collapsed objects with $T_{\rm vir} < 10^4
\kel$.  This condition sets the upper mass limit for minihalos; the
lower mass limit is determined by the Jeans mass or, more precisely,
the time-averaged Jeans mass $M_{\rm fil}$ \citep{gnedin} using the
IGM temperature given in equation (\ref{eq:Tigm}).  If molecular
hydrogen is present, cooling becomes more effective (Ricotti et
al. 2002), but most models predict that even a small UV background
will efficiently dissociate H$_2$ in these halos
\citep{haiman97,ciardi,haiman00}.

We compute the optical depth $\tau_\nu$ of the 21 cm transition along
a line of sight through a minihalo with impact parameter $\alpha$.
For the remainder of this section, $\nu$ refers to the frequency of
the background radiation at the position of the minihalo (i.e., not
including the cosmological redshift between the observer and the
minihalo).  Each collapsed object is surrounded by an accretion shock.
Inside the shock, we assume that the gas overdensity equals that of
the dark matter, which we take to have the universal NFW profile of
\citet{nfw}.  In reality, gas pressure will flatten the core of the
gas profile relative to the dark matter, but we find that lines of
sight passing through the affected regions are so rare that they make
only a negligible difference to our final results.

Outside of the virialization shock, each halo is surrounded by a
region of infalling gas.  In an Einstein-de Sitter universe, the
structure of the infall region can be described by a self-similar
solution \citep{bert}.  At the redshifts we consider ($6 \la z \la
10$), deviations from an Einstein-de Sitter universe are small, and
the solution is approximately valid.  The \citet{bert} profile fixes
the shock to be at a constant overdensity relative to the background
medium: the shock radius $r_{\rm sh} \approx 4 r_{\rm vir}/3$, with a
slight dependence on cosmology and redshift.  The virial radius
$r_{\rm vir}$ is defined in equation (24) of \citet{barkana01}.
Similarly, the turnaround radius (i.e., the radius at which the gas is
at rest with respect to halo) is $r_{\rm ta} \approx 3.85 r_{\rm vir}$.  We
scale the velocity field to match the Hubble flow and the gas density
to match the mean IGM density at large radii.  Unfortunately, the
\citet{bert} profile has a much steeper central cusp ($\rho \propto
r^{-2.25}$) than the NFW profile.  We therefore continue to use an NFW
profile within the accretion shock in order to avoid seriously
overcounting the number of high-optical depth lines of sight.  As a
result, the density jump condition at the shock is only satisfied to
$\sim 10\%$ for the halos we study.  However, because all of the
quantities of interest are integrated along the line of sight, this
error makes little difference to our results.

The HI atoms are assumed to have a Maxwellian distribution at each
point.  For $r \leq r_{\rm sh}$ we set $T_K=T_{\rm vir}$, while for $r
> r_{\rm sh}$, we set $T_K=T_{\rm IGM}$.  Note that we therefore
ignore temperature variations within the shocked region.  The
\citet{bert} solution gives a postshock temperature $T_{\rm sh} \sim
T_{\rm vir}/3$ and thus slightly increases the optical depth. We also
ignore the heating of the gas as it is compressed in the infall
region; in the adiabatic limit, this leads to at most a factor $\sim
2.5$ error in the temperature.  However, because the overdensity in
the infall region is not large, the effects on the total optical
depths are small.  The spin temperature is then determined locally via
equation (\ref{eq:hItspin}).

Once the halo profile is specified, the optical depth at a frequency
$\nu$ along a line of sight with impact parameter $\alpha$ from the
halo center is 
\bq 
\tau_\nu = \frac{ 3 h_P c^3
A_{10}}{32 \pi k \nu_0^2} \int_{-\infty}^\infty dR
\frac{n_{HI}(r)}{T_s(r) \sqrt{\pi} b(r)} \exp \left\{ - \frac{[v(\nu)
- v_{\rm LOS}(\alpha,R)]^2}{b^2(r)} \right\}.
\label{eq:tauhalo}
\eq 
Here, $r^2 = \alpha^2 + R^2$, $v(\nu) = c(\nu-\nu_0)/\nu_0$,
$v_{\rm LOS}(\alpha,R)$ is the infall velocity projected along the
line of sight, and $b^2(r) = 2 k T_K(r)/m_p$ is the Doppler parameter
of the gas.  For $\nu$ sufficiently far from $\nu_0$, the exponential
factor is non-zero only in the diffuse IGM where $n_{HI}$, $T_K$, and
$T_S$ are constants.  At these frequencies, the integral becomes
$n_{HI}/[T_S H(z)]$ due to the Hubble flow, and we recover equation
(\ref{eq:tauigm}).

\begin{figure}[t]
\plotone{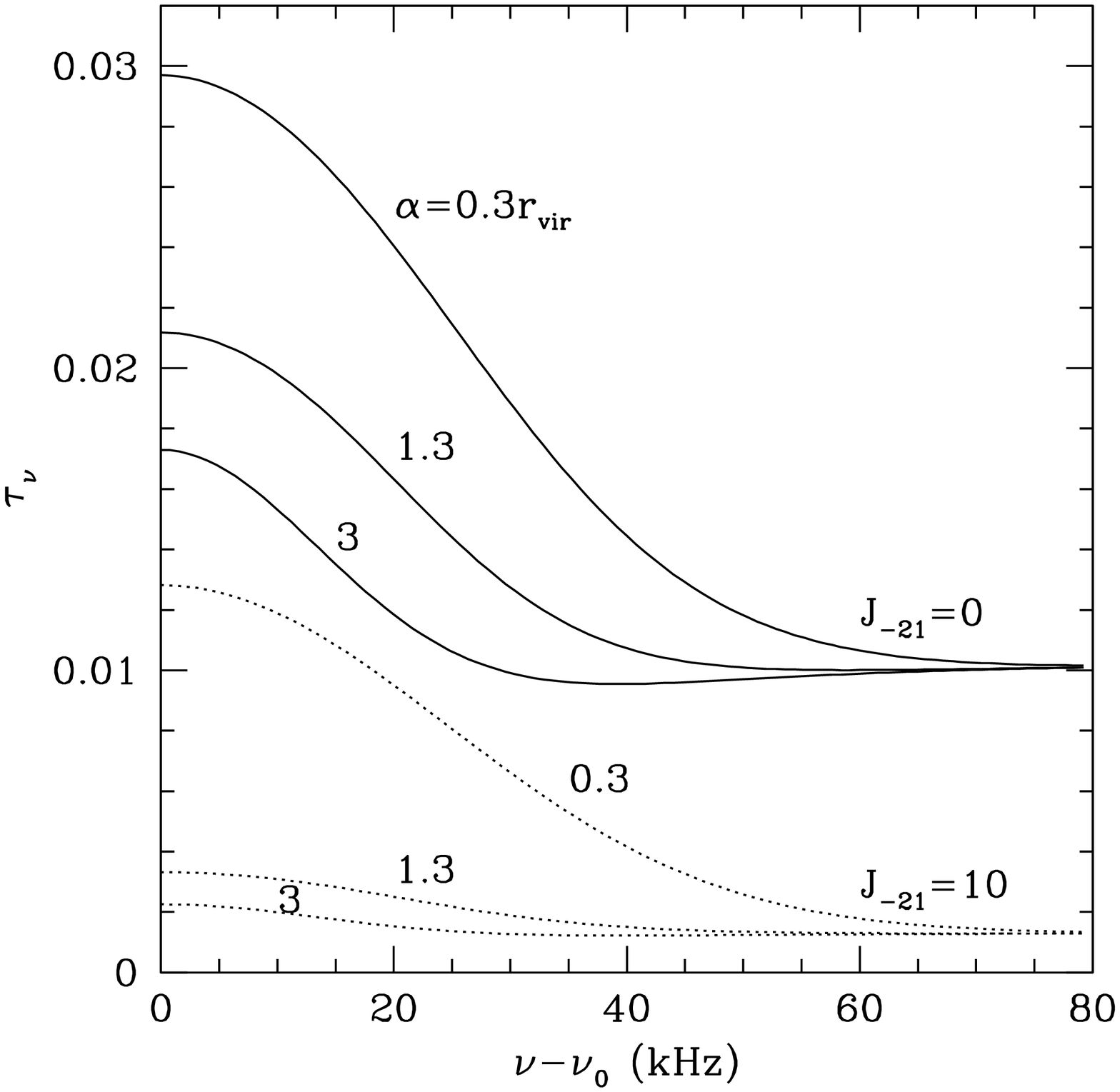}
\caption{ Optical depth profiles for a $M_h=5 \times 10^6 \msun$
minihalo at $z_h=10$ embedded in a heated IGM.  Each set of profiles
assumes $\alpha = 0.3,\,1.3,$ and $3 r_{\rm vir}$, from top to bottom.
The solid curves assume a heated IGM and $J_{-21} = 0$, while the
dotted lines show results for $J_{-21} = 10$. }
\label{fig:taunu}
\end{figure}

Sample optical depth profiles for a $M_h=5 \times 10^6 \msun$ minihalo
at $z_h=10$ are shown in Figure \ref{fig:taunu}.  The solid curves
assume a heated IGM and $J_{-21} = 0$ while the dotted lines show
analogous results for $J_{-21} = 10$.  Each set of profiles assumes
$\alpha = 0.3,\,1.3,$ and $3 r_{\rm vir}$, from top to bottom.  In all
cases, the optical depth at large $\nu-\nu_0$ is simply that of the
IGM, which is strongly suppressed for a strong $T_S$--$T_K$ coupling
through a Ly$\alpha$ background.  Careful examination of the $\alpha =
3 r_{\rm vir}$ curves reveals that $\tau_\nu < \tau_{\rm IGM}$ for
moderately large $\nu-\nu_0$.  This occurs because the velocity
gradients within the infall region exceed that of the Hubble flow, and
so for a range of frequencies the absorbing column is actually smaller
than that of the diffuse IGM.

As is evident from Figure \ref{fig:taunu}, the line width varies with
impact parameter.  For $\alpha \la r_{\rm vir}$, the dominant
contribution to $\tau_\nu$ comes from shocked gas within the halo,
and once absorption due to the diffuse IGM is removed, 
the line profile is close to a Gaussian with a width determined by
$T_{\rm vir}$.  For $\alpha \gg r_{\rm vir}$, the line of sight does not
pass through shocked gas and so the profile is approximately a
Gaussian of width $T_{\rm IGM}$.  For $\alpha \ga r_{\rm vir}$, the
line of sight passes close to the shock, and the profile is best fit
by an effective temperature $T_{\rm eff} = 0.5 (T_{\rm IGM} + T_{\rm
vir}/3)$.  [The factor of $1/3$ results from the infall velocities at
the shock in the \citet{bert} profile; see the paragraph above
equation (\ref{eq:tauhalo}).]  The observed FWHM (defined relative to
the diffuse IGM absorption) is then 
\bq 
\Delta \nu_{\rm obs} = 3.2
\left( \frac{T_{\rm eff}}{10^3 \kel} \right)^{1/2} \left(
\frac{1+z_h}{10} \right)^{-1} \khz,
\label{eq:obswidth}
\eq
where $T_{\rm eff}$ is chosen as described above and $z_h$ is the
redshift of the minihalo.  Note that Figure \ref{fig:taunu} shows
\emph{intrinsic} line profiles (i.e., without the cosmological
redshift between the observer and the halo).

Another quantity of interest is a measure of the total absorption due
to the halo, or the ``equivalent width.''  We define the
\emph{intrinsic} equivalent width through the relation 
\bq
\frac{\langle \Delta \nu \rangle_{\rm int}}{2} = \int_{\nu_0}^{\infty}
(1 - e^{-\tau_\nu}) d \nu - \int_{\nu_0}^{\infty} (1 - e^{-\tau_{\rm
IGM}}) d \nu,
\label{eq:ewdefn}
\eq 
where $\tau_{\rm IGM}$ is given by equation (\ref{eq:tauigm}).
This measures the amount of flux absorbed by the minihalo
over and above that absorbed by the diffuse IGM.  Because of the
cosmological redshift, the \emph{observed} equivalent width $\langle
\Delta \nu \rangle_{\rm obs} = \langle \Delta \nu \rangle_{\rm
int}/(1+z_h)$. 

\begin{figure}[t]
\plotone{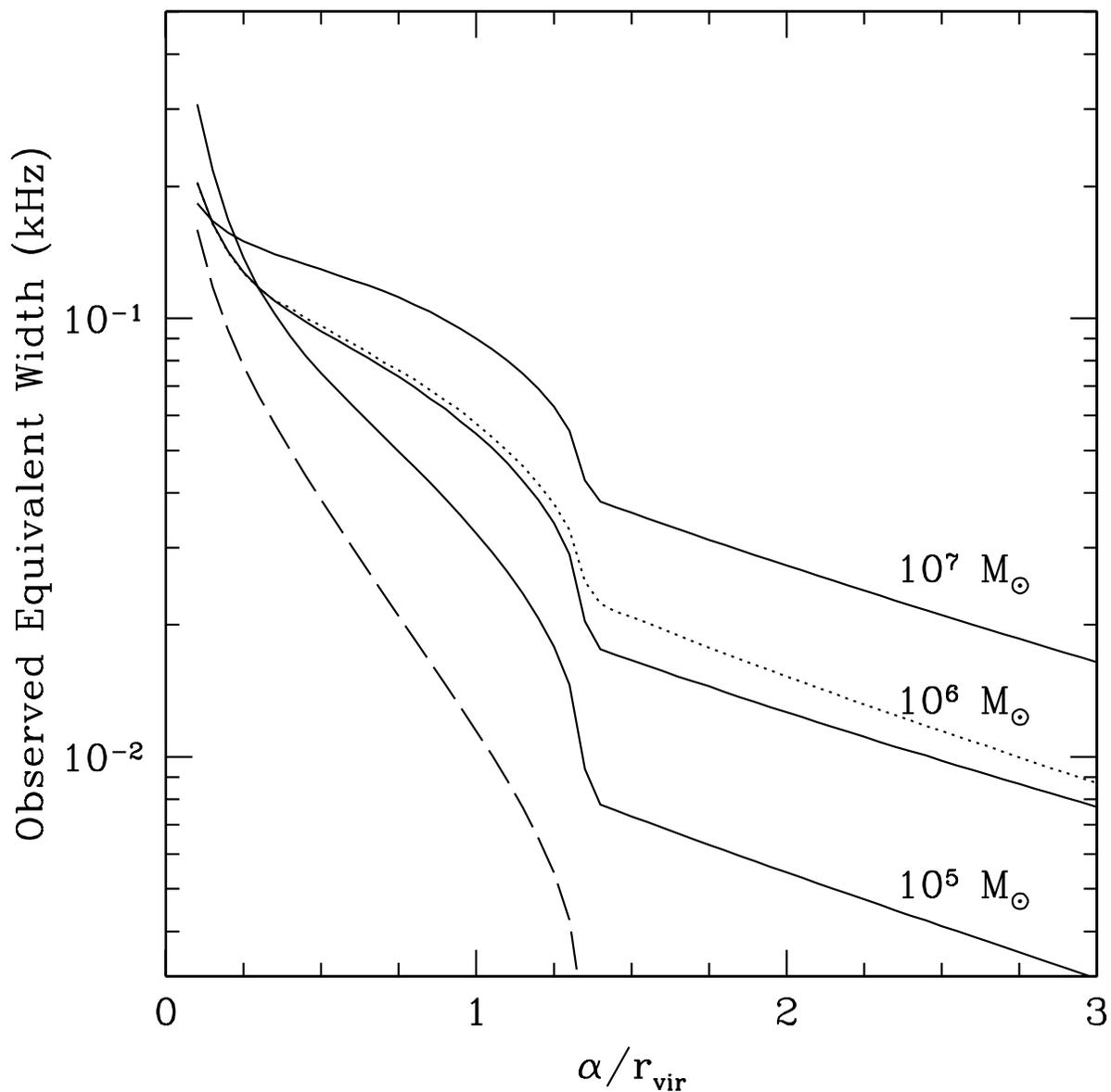}
\caption{ Observed equivalent width as a function of impact parameter
$\alpha$.  Solid curves show results for $M_h=10^7,\,10^6,$ and $10^5
\msun$ from top to bottom, with $J_{-21} = 0$ and a heated IGM. The
dotted curve shows results for $M_h = 10^6 \msun$ and no IGM heating,
while the dashed curve shows results for the same mass, IGM heating,
and a radiation background $J_{-21}=10$.  All curves assume $z_h=8$.  }
\label{fig:ewrad}
\end{figure}

The observed equivalent width as a function of impact parameter is
shown for several $z_h=8$ halos in Figure \ref{fig:ewrad}.  Solid
curves show results for $M_h=10^7,\,10^6,$ and $10^5 \msun$ from top
to bottom , with $J_{-21} = 0$ and a heated IGM. The dotted line shows
results for $M_h = 10^6 \msun$ and no IGM heating, while the dashed
line shows results for the same mass, full IGM heating, and a
radiation background $J_{-21}=10$.  The steep drop at $\alpha/r_{\rm
vir} \approx 1.3$ occurs because of the density drop past the
accretion shock.  Note that the IGM temperature has little effect on
the equivalent width.  \emph{Smaller} halos have \emph{larger} central
equivalent widths (because such halos are cooler, and $\tau_\nu
\propto T_S^{-1} T_K^{-1/2}$) but \emph{smaller} equivalent widths at
the most probable impact parameters because their column density is
smaller.  A cold IGM slightly increases the equivalent width at large
$\alpha$ because more of the absorbing gas is located within the
infall region of the halo, where the density is above the mean IGM
density.  On the other hand, the radiation background has a pronounced
effect on the equivalent widths, particularly at large $\alpha$.  By
effectively coupling $T_S$ to $T_{\rm vir}$ or $T_{\rm IGM}$ in the
outskirts of the halo where collisions are ineffective, such a
radiation field drastically decreases the optical depth contributed by
these regions.

\subsection{ Galactic Disks }

Halos with $T_{\rm vir} > 10^4 \kel$ can cool through atomic line
radiation and are expected to collapse and form protogalactic objects.
We assume here that these objects form disks.  We adopt the
exponential disk structure of \citet{mmw}, in which
\begin{eqnarray}
\Sigma(\alpha) = \frac{ M_d }{2 \pi R_d^2} e^{-\alpha/R_d},
\label{eq:coldisk}
\\
R_d = \frac{1}{\sqrt{2}} j_d \lambda r_{\rm vir}.
\label{eq:rdisk}
\end{eqnarray}
Here $\Sigma$ is the surface density of the disk, $R_d$ is the disk
scale length, $M_d$ is the disk mass, $j_d$ is the specific angular
momentum of the baryons in units of their initial value upon halo
collapse, and $\lambda$ is the spin parameter.  \citet{mmw} found that
$j_d=1$ was required in order to match observational data (implying no
substantial loss of baryonic angular momentum to the dark matter),
given the fact that N-body simulations show that $\lambda$ has a
lognormal distribution with a mean $\bar{\lambda}=0.05$.  We will
assume for simplicity that all disks have this mean spin parameter
${\bar \lambda}$.

The primary contribution to HI absorption in galaxies comes from the
cold neutral medium.  We assume that a fraction $f_d$ of the total
baryonic mass of the halo, $(\Omega_b/\Omega_m) M_h$, is in this
phase.  (The remainder can be in stars, the hot interstellar medium,
or a hot component of the halo.)  We assume for simplicity that the
cold gas is smoothly distributed throughout the disk (although in
reality it is likely to be bound up in small, dense clouds dispersed
throughout the disk).  We take a spin temperature $T_S = 10^3 \kel$,
as inferred from observations of damped Ly$\alpha$ absorbers at lower
redshifts \citep{carilli96}.  We further assume that the spin and
kinetic temperatures are coupled, as expected for the radiation
background in a typical galaxy.  The optical depth is then 
\bq
\tau_\nu = \tau_{\rm IGM} + \tau_0 \exp
\left[-\frac{v^2(\nu)}{b^2(T_K)} \right],
\label{eq:taudisk}
\eq
where $\tau_0 = \tau_0^c e^{-\alpha/R_d}$ and 
\begin{eqnarray}
\tau_0^c & = & 15.1 \, h^{4/3} \left( \frac{f_d}{0.5} \right) \left(
\frac{\Omega_b}{0.05} \right) \left( \frac{\Omega_m}{0.3} \right)^{-1/3}
\left( \frac{T_S}{10^3 \kel} \right)^{-3/2} \nonumber \\
\, & \, & \times \left(\frac{\bar{\lambda}}{0.05} \right)^{-2} \left(
\frac{M_h}{10^8 \msun} \right)^{1/3} \left( \frac{1+z}{10} \right)^2.
\label{eq:tauc}
\end{eqnarray}
In this expression, we have assumed a random distribution of disk
inclinations and have taken the average path length through an inclined
disk (so that $\Sigma_0 \rightarrow \pi^2/4 \Sigma_0$).  We have
ignored the modifications to $\tau_{\rm IGM}$ caused by the velocity
gradient and overdensity of the infall region, which differs slightly
from the Hubble flow (and therefore, at a given velocity offset, the
column of neutral hydrogen differs from that of the unperturbed IGM).
Typically, the large optical depths of lines of sight through galaxies
make such subtle differences unimportant.  However, note that disks
are likely to be in complicated environments, both because of feedback
from the collapsed object and because high-redshift protogalaxies are
expected to be highly clustered.  Feedback can heat and ionize the
region around the disk, reducing the IGM absorption well below the
nominal $\tau_{\rm IGM}$.  The equivalent width can be calculated from
equation~(\ref{eq:ewdefn}); with our assumption of a constant
$T_S=T_K$ within each system, the relation becomes  
\bq
\langle \Delta \nu \rangle_{\rm int} = 34 \khz \, \left(
\frac{T_S}{10^3 \kel} \right)^{1/2} \sum_{n=1}^\infty \frac{(-1)^{n-1}
\, \tau_0^n}{n! \, \sqrt{n}}.
\label{eq:ewtaudisk}
\eq

\section{ Results }
\label{results}

\subsection{ Number Densities }

Using our model we can compute the density of lines of a given peak
optical depth or equivalent width in a spectrum.  For a quantity
$\zeta$ (either $\tau_0$ or equivalent width), the number of systems
intersected with $\zeta > \zeta_0$ per redshift interval is 
\bq
\frac{d N(>\zeta_0)}{d z} = (1+z)^2 \, \frac{d r}{d z} \, \int_{M_{\rm
min}}^{M_{\rm max}} d M_h \, \frac{d n_h}{d M_h} A(M_h,z,\zeta_0).
\label{eq:dndzdefn}
\eq 
Here $dr/dz$ is the comoving length element per unit redshift, $d
n_h/d M_h$ is the halo mass function (or number of halos at redshift
$z$ with mass between $M_h$ and $M_h+dM_h$ per comoving volume), and
$A(M_h,z,\zeta_0)$ is the cross-sectional area of a halo (in physical
coordinates) with the specified mass and redshift subtended by lines
of sight for which $\zeta > \zeta_0$.  For minihalos, $M_{\rm min} =
M_{\rm fil}$ and $M_{\rm max}$ is determined by the cooling criterion.
For disks, $M_{\rm min}$ is determined by cooling and $M_{\rm
max}=\infty$.  In both cases, we use $dn_h/dM_h$ given by the
Press-Schechter formalism \citep{press}, with the modifications
suggested by \citet{jenkins} in order to better match the mass
function found by numerical simulations.  For disk calculations, we
again assume a random distribution of disk inclinations; the effective
area of each disk is therefore $4/\pi^2$ times the area of the disk
viewed face-on.  In this case the area is simply 
\bq
A(M_h,z,\tau_0) = \frac{2 \bar{\lambda}^2 r_{\rm vir}^2}{\pi} \, \ln^2
\left( \frac{\tau_0^c(M_h,z)}{\tau_0} \right).
\label{eq:areadisk}
\eq

\begin{figure}[t]
\plotone{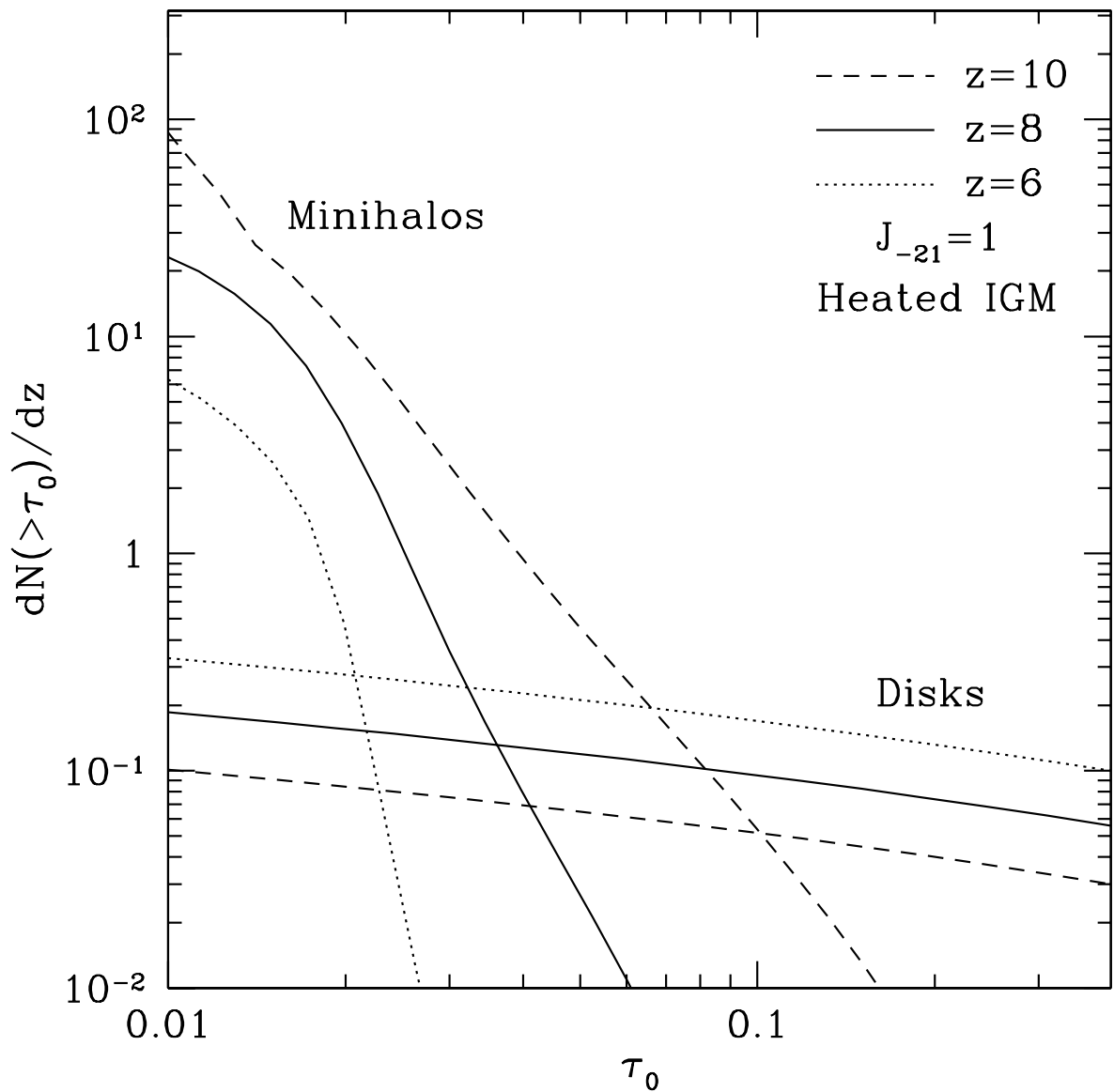}
\caption{ Number of systems with observed central optical depth
greater than $\tau_0$ intersected per redshift interval.  The dashed,
solid, and dotted curves show results for $z=10,\,8,$ and $6$,
respectively.  The two sets of lines describe minihalos and
protogalactic disks.  Minihalo calculations assume $J_{-21}=1$ and a
heated IGM.  }
\label{fig:dndtz}
\end{figure}

Figure \ref{fig:dndtz} shows $dN(>\tau_0)/dz$, the number of objects
intersected per redshift interval with a central optical depth greater
than $\tau_0$.  The dashed, solid, and dotted curves show results for
$z=10,\,8,$ and $6$, respectively.  All minihalo calculations assume a
heated IGM and $J_{-21}=1$.  The density of minihalo lines decreases
with cosmic time because the minimum halo mass increases as $T_{\rm
IGM}$ increases.  Note that the evolution with redshift is
particularly pronounced for large $\tau_0$.

It is clear that minihalo absorption systems are much more common than
disks at low optical depths, but the rare high-optical depth features
are dominated by collapsed disks.  The flat disk distribution results
from the exponential dependence of the optical depth on impact
parameter.  The primary uncertainty in our disk calculation is $T_S$.
A single cold cloud in the Milky Way typically has $T_S \sim 100 \kel$
\citep{dickey}, but observations of moderate-redshift absorbers
suggest $T_S \sim 10^3 \kel$ \citep{carilli96}.  If $T_S$ is smaller
than we assume, the optical depth through the disk increases
substantially [$\tau_0 \propto T_S^{-3/2}$; see equation
(\ref{eq:tauc})].  However, as equation (\ref{eq:areadisk}) shows, the
distribution varies only logarithmically with $\tau_0^c$, so the
corresponding change in $dN(>\tau_0)/dz$ will not be large.  For
example, in the range of optical depths shown in Figure
\ref{fig:dndtz}, setting $T_S = 10^2 \kel$ increases the number
density of lines by a factor of only $\sim 2$.

\begin{figure}[t]
\plotone{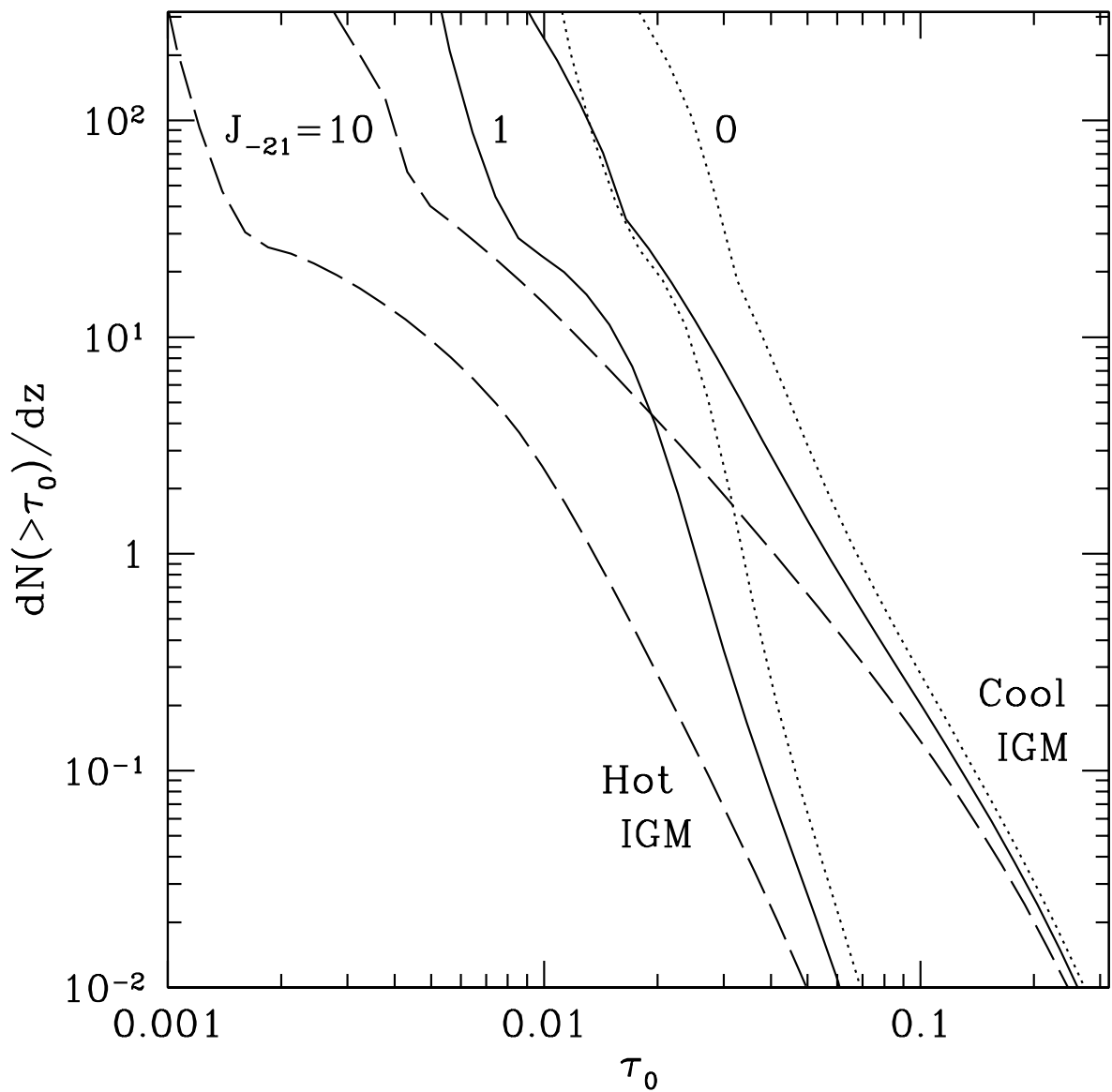}
\caption{ Number of systems with observed central optical depth
greater than $\tau_0$ intersected per redshift interval at $z=8$.
The solid curves assume $J_{-21}=1$, the dotted curves assume
$J_{-21}=0$, and the dashed curves assume $J_{-21}=10$.  Within each
set, the leftmost curve assumes our standard IGM temperature and the
rightmost curve assumes $T_{\rm IGM}$ is an order of magnitude below
that of equation (\ref{eq:Tigm}).  }
\label{fig:dndtj}
\end{figure}

Figure \ref{fig:dndtj} shows the variation of $dN(>\tau_0)/dz$ with
IGM properties at $z=8$.  The solid curves assume $J_{-21}=1$, the
dotted curves assume $J_{-21}=0$, and the dashed curves assume
$J_{-21}=10$.  Within each set, the left line assumes our standard IGM
temperature and the right line assumes that $T_{\rm IGM}$ is an order
of magnitude below the value given by equation (\ref{eq:Tigm}).  The
steep rise as $\tau_0$ approaches $\tau_{\rm IGM}$ is caused by the
divergence of the relevant cross-section for each halo in this limit.
We see that the radiation background can be inferred from the
absorbing continuum while $T_{\rm IGM}$ can be estimated from the
statistics of deep lines.  At small $\tau_0$, the distribution is
primarily determined by the radiation field $J_{-21}$.  On the other
hand, $d N(>\tau_0)/dz$ depends much more strongly on $T_{\rm IGM}$
than on $J_{-21}$ at large $\tau_0$.  Such large optical depths
require a line of sight passing close to the halo center, where the
gas density is high enough for collisional coupling between $T_S$ and
$T_K$ to be effective, even without a substantial radiation field.
The temperature dependence arises because a cool IGM has a smaller
$M_{\rm fil}$ and thus a larger number of low-mass halos with small
$T_{\rm vir}$ (or large $\tau_0$).

\citet{carilli} examined lines of sight through a cosmological
simulation and estimated that $dN(\tau_0 > 0.02,z=10)/dz \sim 50$ and
$dN(\tau_0 > 0.02,z=8)/dz \sim 4$.  For their simulation, the average
radiation background evolves between $J_{-21} \sim 1$--$4$ over this
redshift range.  Figure \ref{fig:dndtj} shows that their $z=8$ result
lies in the range we would predict for this radiation field, but their
$z=10$ point is $\sim 4$ times higher than our prediction in Figure
\ref{fig:dndtz}.  [If we choose $T_{\rm IGM}$ to precisely match the
mass-averaged temperature at $z=10$ in the simulation of
\citet{carilli}, the discrepancy is reduced to a factor of $\sim 3$.]
A difference in results should be expected given that the two
calculations are complementary and do not consider the same population
of absorbers.  The baryonic mass per particle in the simulation was
$10^{5.7} \msun$ \citep{gnedin02}, while even the largest of our
minihalos have baryonic masses $\sim 10^{6.7} \msun$.  Our formalism
therefore probes objects of much smaller mass than does the
simulation; on the other hand, the simulation probes the filamentary
structure of the IGM which is not adequately described by our
formalism.  Furthermore, the simulation includes non-uniform heating
and radiation.  Overall, it appears that filaments and minihalos
provide comparable contributions to the line statistics, with the
relative importance of filaments increasing as redshift increases.

\begin{figure}[t]
\plotone{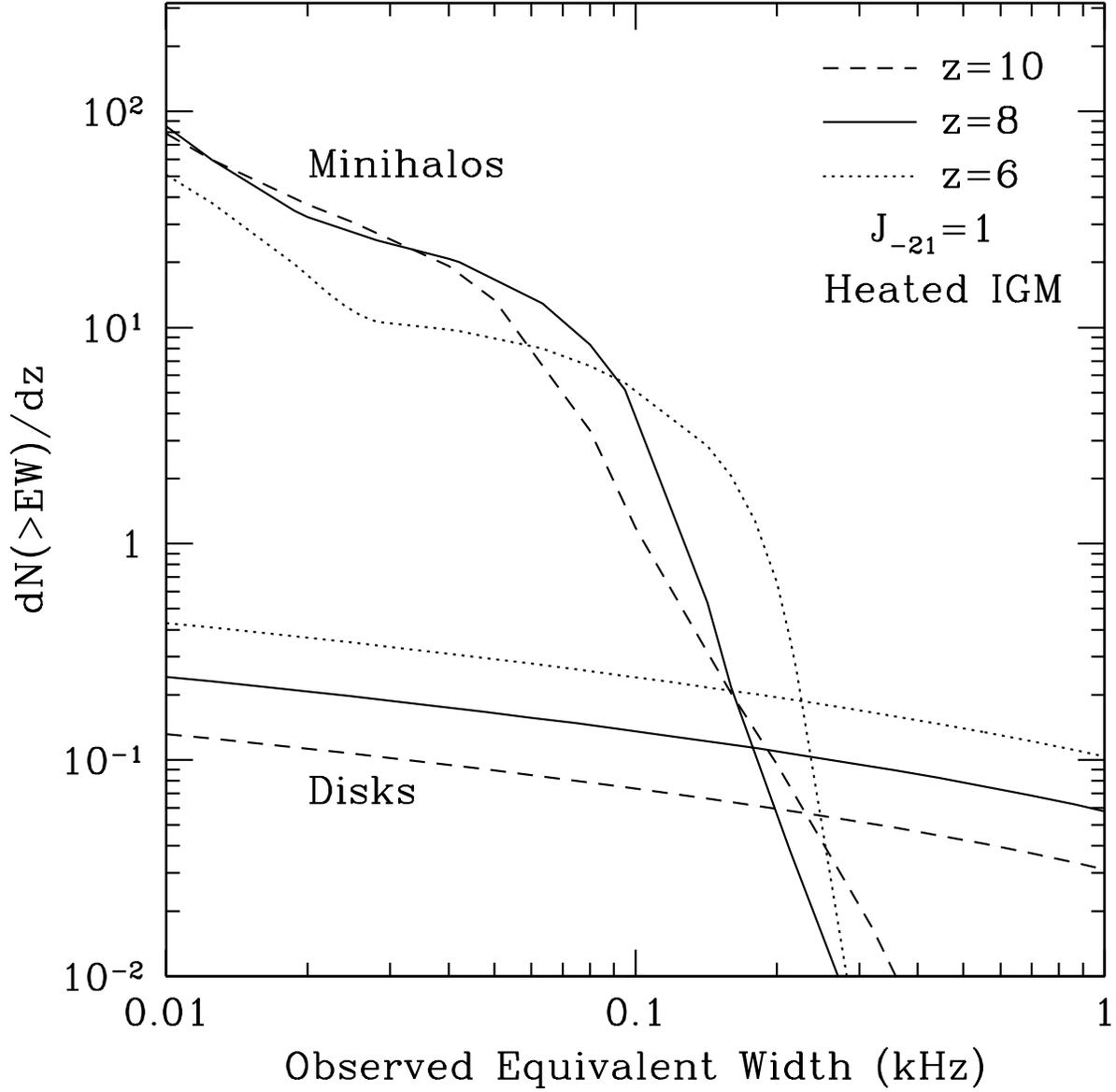}
\caption{ Number of systems with observed equivalent width greater
than a threshold intersected per redshift interval, as a function of
the threshold value.  The dashed, solid, and dotted curves show
results for $z=10,\,8,$ and $6$, respectively.  The two sets of lines
describe minihalos and protogalactic disks.  All minihalo calculations
assume a heated IGM and $J_{-21} =1$. }
\label{fig:dnewz}
\end{figure}

Another useful quantity is the number of intersections per redshift
interval with observed equivalent widths greater than a fixed value,
$dN(>EW)/dz$.  In Figure \ref{fig:dnewz} we show this for the same
redshifts and IGM characteristics as in Figure \ref{fig:dndtz}.
Again, disks only dominate the number counts for very large equivalent
widths.  In contrast to the optical depth statistics, we find little
evolution of $dN(>EW)/dz$ with redshift.  This is caused in part
because our definition of the equivalent width removes the effect of
$\tau_{\rm IGM}$ (which decreases rapidly with cosmic time). In
addition, the small, high optical depth halos at high-$z$
yield narrow lines and therefore do not have substantial equivalent
widths.  As a result, the simultaneous increases in $M_{\rm fil}$ and
minihalo density with cosmic time appear to roughly balance each
other.  We do not show the dependence of the equivalent width
statistics on $J_{-21}$ and $T_{\rm IGM}$ because it exhibits the same
trends as $dN(>\tau_0)/dz$, although at a weaker level.

\subsection{ Synthetic Spectra }

Our model can also be used to generate Monte Carlo realizations of
absorption spectra for illustrative purposes.  The probability $P(z)$
that a line of sight intersects an object in a redshift interval $dz$
centered on $z$ is given by equation (\ref{eq:dndzdefn}), with
$A(M_h,z)$ simply detemined by the maximum impact parameter allowed in
the calculation.  (We find that choosing $\alpha_{\rm max} = 5 r_{\rm
vir}$ is sufficient to include all observable absorption lines from
minihalos.)  To generate a spectrum, we choose redshift intervals
$\Delta z$ small enough so that $P(z) < 0.01$.  For each bin, we
generate a random number to determine if an intersection takes place.
If one occurs, we choose $M_h$ randomly using the probability
distribution $d n_h/d M_h$ and then randomly select the impact
parameter with an equal probability per cross-sectional area of the
halo.  Note that this method does not include clustering of minihalos
and in principle allows overlap between minihalos centered in nearby
bins (although in practice such overlap is highly insignificant).

\begin{figure}[t]
\plotone{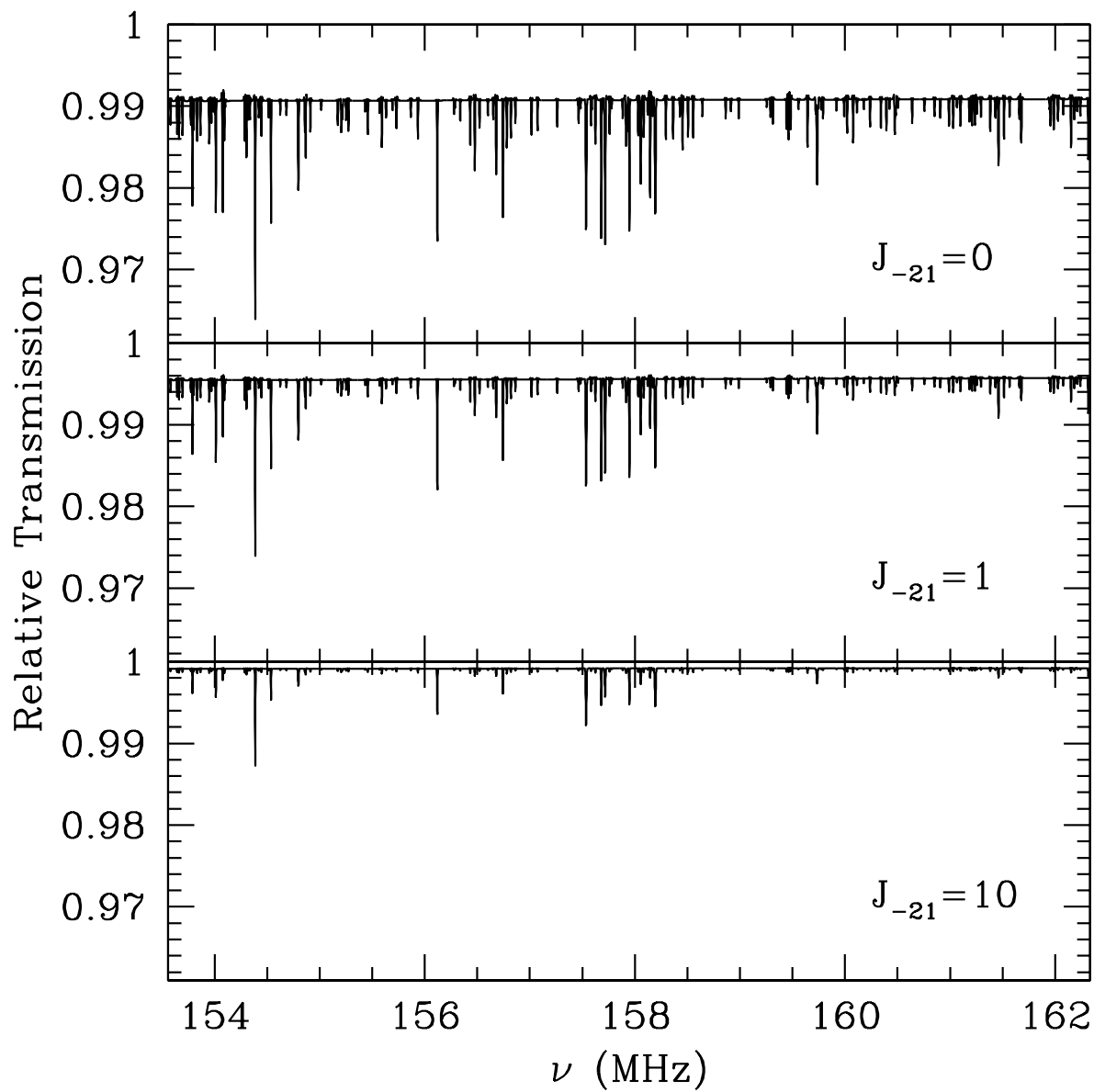}
\caption{ Simulated transmission along a line of sight to a distant radio
source assuming absorption by intervening minihalos.  All panels are
generated using the same realization of the minihalo mass field and assume
a heated IGM.  The panels show $J_{-21}=0,\,1,$ and $10$, from top to
bottom. }
\label{fig:mcspec}
\end{figure}

Sample transmission curves generated in this way and smoothed to $1
\khz$ resolution are shown in Figure \ref{fig:mcspec}.  Here the
transmission $T$ is defined as $T = 1 - \exp(-\tau_\nu)$.  In each
panel, we show the portion of a spectrum from $z=8.25$ to $z=7.75$.
Each panel is generated from the same set of intersected halos, but
assumes a different radiation field in the calculation of the optical
depths ($J_{-21}=0,\,1,$ and $10$, from top to bottom).  The
figure explicitly shows that by coupling $T_S$ and $T_K$, a radiation
field can strongly decrease both the absorption of the diffuse IGM and
of the minihalos.  [For reference, $T_{\rm IGM}(z=8) = 3162 \kel$ in
our model.]  The strongest absorption is caused by a $2 \times 10^6
\msun$ minihalo at $z=8.2$, with $\alpha = 0.35 r_{\rm vir}$.  Close
inspection reveals that most absorption lines are surrounded by a
small region with transmission greater than that of the diffuse IGM.
Such features result from the steep velocity gradients in the infall
region of a halo, which cause the optical depth to fall below
$\tau_{\rm IGM}$ by small factor (see Figure \ref{fig:taunu}).

Similarly, one can generate realizations of spectra including disk
intersections.  Because these intersections are rare (see Figure
\ref{fig:dndtz}), we performed a Monte Carlo procedure on the interval
between $z=8.25$ and $z=7.75$ sixteen times in order to find a line of
sight passing through a disk with $\tau_0 > 0.1$.  [Figure
\ref{fig:dndtz} shows that for disks, $dN(\tau_0>0.1)/dz \approx 0.09$
at $z=8$.  On average, we would therefore expect one such intersection
per $22$ realizations of this redshift interval.]  In fact, the
resulting system has $\tau_0 = 3.76$.  We then added this system to an
independently generated minihalo field, and we show the resulting
spectrum in Figure 8.  In order to contrast the effects of disk and
minihalo intersections, we switch to logarithmic axes and plot the
flux decrement, $1-T=\exp(-\tau_\nu)$, instead of the transmission.
The strong feature is caused by the disk, while the much weaker
features to the right are caused by minihalos.  Although such lines of
sight are obviously rare, we see that disks can in general be clearly
distinguished from the minihalo population because they have much
larger peak optical depths and equivalent widths.  Note that the line
width here is a direct result of our assumption that $T_K=10^3 \kel$.
Most damped Ly$\alpha$ absorbers at lower redshifts show more than one
velocity component of somewhat narrower intrinsic width
\citep{briggs,lane}.

\section{ Discussion }
\label{disc}

We have shown that absorption at the 21 cm transition of neutral
hydrogen in both minihalos and protogalactic disks at high-redshifts
can cause non-negligible absorption in the low-frequency spectra of
high-redshift radio sources.  Using a semi-analytic approach, we have
found that the density of minihalo absorption features is primarily a
function of the kinetic temperature of the IGM (which determines the
minimum mass of the minihalos) and the radiation background (which
couples the spin and kinetic temperatures of the hydrogen gas).  We
predict that the average radiation background can be measured by the
continuum absorption and the number density of very weak lines, while
the IGM temperature can be measured by the density of narrow,
relatively strong absorption lines.  Optical depths reach $\tau_0 \sim
0.02$ for objects common enough to appear along typical lines of
sight.  Studies of absorption radio spectra therefore offer a probe of
the era before reionization and can help constrain the history of
early star formation.  We find, unsurprisingly, that disks are only
rarely intersected but tend to have much higher optical depths.

The principal shortcoming of our semi-analytic approach is its
inability to accurately describe the filamentary structure of the IGM
and the spatial clustering of halos.  Using a simulation,
\citet{carilli} have also studied 21 cm absorption along lines of
sight to high-redshift radio sources.  While the mass resolution of
their simulation was too coarse to include the minihalo population,
they were able to capture the geometry and structure of filaments.
They find that both the typical optical depths and the number
densities of filament absorption lines are comparable to or slightly
larger than our estimates for those from minihalos.  An observed
absorption spectrum will then be a mixture of lines from these two
populations of absorbers.  The clustering of minihalo lines may yield
information on bias or feedback from nearby cooled objects, but a
proper study of such an effect will require numerical simulations or
more detailed semi-analytic models (e.g., \citealt{scann}).

Can these signals be detected by future radio telescopes?  First, note
that the relevant frequency range is $100-200 \mhz$, corresponding to
$z \sim 13$--$6$; thus, a low-frequency instrument is necessary.  The
transmission curves in \S 3.2 ignore noise.  In a real observation,
the noise level will be determined by a combination of the telescope
characteristics and the brightness of the background source.  Because
the specifications of the next generation of telescopes are not yet
settled, we have chosen not to present results specifically tailored
to any particular instrument.  Instead, we can simply estimate the
required source brightness, $S_{\rm min}$, in order to observe a
single absorption feature, given the telescope's effective area
$A_{\rm eff}$, system temperature $T_{\rm sys}$, and channel width
$\Delta \nu_{\rm ch}$, for a specified signal-to-noise ratio $S/N$ and
integration time $t$.  In the small $\tau_0$ limit and under the
assumption that $\Delta \nu_{\rm ch} \ll \Delta \nu_{\rm obs}$, the
result is 
\bq 
S_{\rm min} = 12 \mjy \, \left( \frac{S/N}{5} \right)
\left( \frac{0.02}{\tau_0} \right) \left( \frac{ 2 \times 10^3
\aefftsys}{A_{\rm eff}/T_{\rm sys}} \right) \left( \frac{10 \days}{t}
\right)^{1/2} \left( \frac{\khz}{\Delta \nu_{\rm ch}} \right)^{1/2}.
\label{eq:sflux}
\eq 
Current designs for the SKA call for $A_{\rm eff}/T_{\rm sys} \sim
2 \times 10^3 \aefftsys$, while the \emph{Low Frequency
Array}\footnote{See http://www.lofar.org/index.html.} (LOFAR) is
expected to have $A_{\rm eff}/T_{\rm sys} \sim 4.5 \times 10^2
\aefftsys$.  A channel width of $1 \khz$ should be achievable with
LOFAR while maintaining a total spectral coverage $\sim 4 \mhz$ and
with SKA while maintaining a total spectral coverage of $\sim 10
\mhz$.

Do objects of this brightness exist at sufficiently high redshifts?
\citet{haiman01} have shown that, in the context of hierarchical cold
dark matter models, black holes with masses $\ga 10^8 \msun$ can
plausibly exist even beyond $z \sim 10$.  
While we currently have no
observational constraints on this model, \citet{carilli} have also
presented a plausibility argument for the existence of luminous
radio-loud quasars at $z \ga 7$.  By extrapolating the radio galaxy
luminosity function to high redshifts using the observed decline of
optical quasars, they estimate $\ga 2000$ sources on the sky of
sufficient brightness to be useful.  Furthermore, they point out that
a bright radio galaxy at $z=5.2$ has already been detected
\citep{breugel}, and there is no obvious environmental reason for a
high-$z$ cutoff.

Because of their high optical depths, disks could be detected along
lines of sight to much fainter sources.  In this case, equation
(\ref{eq:sflux}) must be modified to apply to large optical depths
through the substitution $\tau_0 \rightarrow 1-\exp(-\tau_0)$.  For
example, detecting a disk with $\tau_0=1$ given the telescope
parameters in equation (\ref{eq:sflux}) would require an average
source flux $\sim 0.4 \mjy$.  Extreme starburst galaxies and faint
active galactic nuclei could satisfy this criterion and of course will
be more common than the bright quasars useful for minihalo studies.
However, the disk intersection probability is sufficiently small that
only a small fraction of all lines of sight will pass through
intervening disks, so searches using these sources are unlikely to be
practical given the long integration times required.  Another
particularly interesting type of source is a gamma-ray burst (GRB).
In this case, the rarity of disks is not a limit because the host
galaxy is necessarily present at the redshift of the burst.  (While
the host galaxy of a faint AGN or a starburst may also be visible, the
gas is much more likely to be ionized on large scales around such
sources.)  The smooth power-law spectrum of GRBs is an ideal continuum
for these observations.  The typical peak flux of a GRB is $\la 1
\mjy$ \citep{ciardigrb}; however, the low-frequency spectra ($\la 10
\ghz$) of GRBs typically suffer from synchrotron self-absorption
\citep{piran}, reducing the flux at $1420 \mhz$ by a factor of a few
from the peak value (e.g., \citealt{galama}).  GRBs are therefore too
dim for the study of minihalo absorption features, but absorption
by the host disk should be observable provided that $\tau_0 \ga 2$.
Such a detection could confirm the redshift of the burst and provide
information on the neutral hydrogen content of its host galaxy.

\acknowledgements

We thank B. Gaensler for helpful discussions.  This work was supported in
part by NASA grants NAG 5-7039, 5-7768, and NSF grants AST-9900877,
AST-0071019 for AL.  SRF acknowledges the support of an NSF graduate
fellowship.

\begin{figure}[t]
\plotone{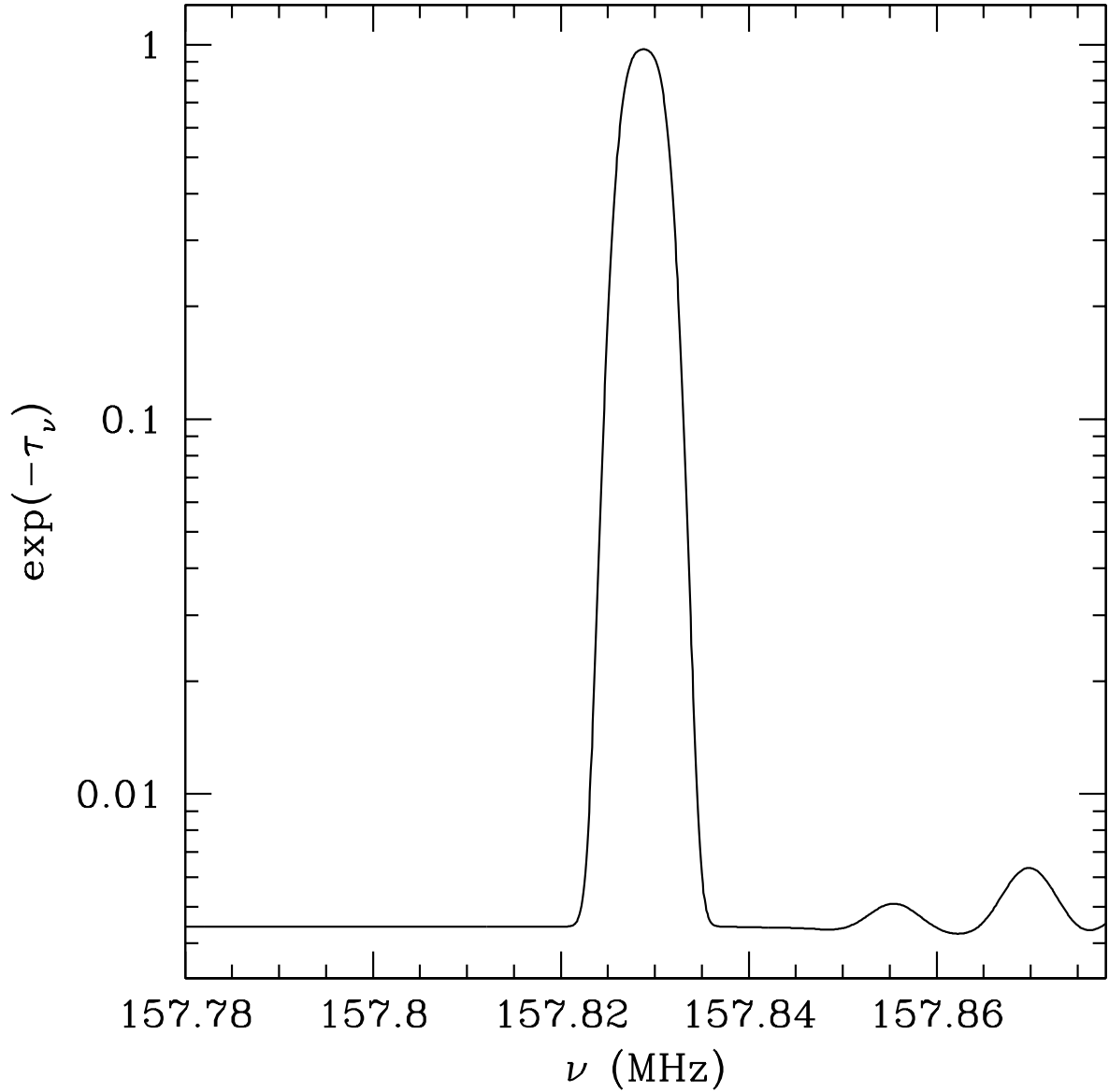}
\caption{  Simulated flux decrement profile along a sightline to a
distant radio source chosen to intersect a protogalactic disk.  The
accompanying minihalo mass field assumes a heated IGM and
$J_{-21}=1$. }
\label{fig:mcdisk}
\end{figure}

\begin{comment}
\end{comment}
\end{document}